\documentclass[
  aps,
  prl,
  twoside,
  twocolumn,
  floatfix
]{revtex4}

\usepackage{amsmath}
\usepackage{amssymb}
\usepackage{graphicx}
\usepackage{tabularx}
\usepackage{dcolumn}
\usepackage{longtable}
\usepackage{footnote}
\usepackage{appendix}

\begin{document}

\newcommand{\bec}{\begin{center}}
\newcommand{\ec}{\end{center}}
\newcommand{\be}{\begin{equation}}
\newcommand{\ee}{\end{equation}}
\newcommand{\beqn}{\begin{eqnarray}}
\newcommand{\eeqn}{\end{eqnarray}}
\newcommand{\bet}{\begin{table}}
\newcommand{\ent}{\end{table}}
\newcommand{\bib}{\bibitem}

\newcommand{\sect}[1]{Sect.~\ref{#1}}
\newcommand{\fig}[1]{Fig.~\ref{#1}}
\newcommand{\Eq}[1]{Eq.~(\ref{#1})}
\newcommand{\eq}[1]{(\ref{#1})}
\newcommand{\tab}[1]{Table~\ref{#1}}

\renewcommand{\vec}[1]{\ensuremath\boldsymbol{#1}}
\renewcommand{\epsilon}[0]{\varepsilon}

\newcommand{\cmt}[1]{\emph{\color{red}#1}%
  \marginpar{{\color{red}\bfseries $!!$}}}



\title{
The classical molecular dynamics simulation
of graphene on Ru(0001)
using a fitted Tersoff interface potential
}


\author{P. S\"ule, M. Szendr\H o} \email{sule@mfa.kfki.hu}

\affiliation
{Research Centre for Natural Sciences,
Institute for Technical Physics and Materials Science
\\
Konkoly Thege u. 29-33, Budapest, Hungary,sule@mfa.kfki.hu,
* The University of E\"otv\"os L\'or\'and, Department of
Materials Physics, Budapest.
\\
}

\date{\today}


\begin{abstract}
The accurate molecular dynamics simulation of weakly bound
adhesive complexes, such as supported graphene, is challenging due to the lack of
an adequate interface potential.
Instead of the widely used Lennard-Jones potential for weak and long-range
interactions 
we use a newly parameterized Tersoff-potential for graphene/Ru(0001)
system.
The new interfacial force field provides adequate moir$\acute{e}$ superstructures
in accordance with scanning tunnelling microscopy images and with
DFT results.
In particular, the corrugation of $\xi \approx 
1.0 \pm 0.2$ $\hbox{\AA}$ is found which is somewhat smaller than found
by DFT approaches ($\xi \approx
1.2$ $\hbox{\AA}$) and is close to STM measurements ($\xi \approx
0.8 \pm 0.3$ $\hbox{\AA}$). 
The new potential could open the way towards large scale simulations
of supported graphene with adequate moir$\acute{e}$ supercells
in many fields of graphene research.
Moreover, the new interface potential might provide a new strategy
in general for getting accurate interaction potentials for weakly
bound adhesion in large scale systems in which atomic dynamics is inaccessible
yet by accurate DFT calculations.
\\
{\em keywords}: atomistic and nanoscale simulations, molecular dynamics simulations, corrugation of graphene, moire superstructures
\end{abstract}

\maketitle

\section{Introduction}

 Since the reliable {\em ab initio}   
density functional theory (DFT) methods with various van der Waals (VDW) functionals
can be used only up to $\sim 1000$ atoms \cite{DFTgr}
 the development of an adequate classical interfacial force-field
for supported graphene (gr) is crucial.
 Although, classical molecular dynamics (CMD) simulations can not be used for the study
of electronic structure, however, many important properties of
graphene, such as surface topography, moire supercells and interfacial
binding characteristics, adhesion as well as defects and many other features can in principle
be studied by CMD simulations \cite{airebo,CMD} without the
explicit consideration of the electron density or orbitals.

The size-limit of DFT
is especially serious if accurate geometry optimization
or molecular dynamics simulations
should be carried out.
Moreover, various VDW DFT functionals provide diverging results \cite{DFTgr,RPA}.
In particular,
accurate DFT potential energy curves for the interface of gr-substrate
systems has been obtained only by the extremely expensive random
phase approximation (RPA) for small modell systems which  
fail, however, to include the full moire-superstructure \cite{RPA,Batzill,rev}.
Using CMD simulations no size-limit problem occurs.
An important limitation here, however, is the limited availability of accurate
interface potential for weak interactions.
The widely accepted choice is the simple pairwise Lennard-Jones (LJ) potential.
In the present paper we point out that this potential
is inadequate, however, for gr/Ru(0001) due to the improper prediction of the
binding sites (registry).
Therefore, the development of reliable interatomic potentials for gr-support systems
is inevitable.

 First principles calculations (such as density functional theory, DFT) have widely been
used in the last few years to understand corrugation of nanoscale gr sheets
on various substrates \cite{DFT:Ru-Hutter,DFT:Ru-Stradi,Peng,DFT:Ru_Wang,DFT:Ru_corrug},
the modelling of larger systems above $1000$ Carbon atoms
remains, however challenging.
Due to the lack of a reliable interface potential for supported gr
until now, the results of CMD simulations of supported gr layer
have been reported in few cases only for SiC support \cite{gr-md}.
Results are also published very recently for gr/Cu(111)
using CMD simulations using LJ potential \cite{gr-md-cu,gr-md-cu2}.

 We find that simple pair potentials are unable to display the moire-supercells
of various gr/substrate systems.
We will show that
simple pairwise potentials 
favor improper binding sites because of the exclusion of
adequate angular orientations at the interface.
Therefore, a new parameter set has been developed for the angular-dependent
(Tersoff) potential at the interface which describe
gr-Ru(0001) bonding adequately.

 We would like to show that using a newly
parameterized interfacial Tersoff-potential 
one can account for the observed surface reconstructions of gr (moire superstructures).
 The experimentally seen
superstructures and corrugation for gr/Ru(0001) \cite{Batzill,rev} are reproduced for the first time then by classical
molecular dynamics simulations using the new parameterized Tersoff potential
at the C-Ru interface.

\section{Methodology}

 Classical molecular dynamics has been used as implemented
in the LAMMPS code (Large-scale Atomic/Molecular Massively Parallel Simulator) \cite{lammps}.
The graphene  layer has been placed incommensurately on the substrates.
The commensurate displacement is energetically
inaccessible (unfavorable), since on-top positions are available
only for a certain part of Carbon atoms when lattice
mismatch exceeds a critical value ($\sim 2-3$ $\%$) \cite{DFT:Ru-Hutter}.
The proper lateral adjustment of C-C bonds would require too much
strain which is clearly unfavorable for most of the
gr-support systems \cite{DFT:Ru-Hutter}. 
The graphene atoms form then an incommensurate overlayer by occupying
partly registered positions (alternating hexagonal hollow and ontop sites) which
leads to a moir$\acute{e}$ superstructure
(long-wave coincidence structures) \cite{rev}.

 The molecular dynamics simulations allow the optimal lateral
positioning of the gr layer in order to reach nearly epitaxial displacement
and the minimization of lattice misfit.
The relaxation of the systems have been reached in two steps:
first geometry optimization has been carried out and then
CMD simulations have been utilized in
few tens of a thousand simulation steps to allow the further reorganization
of the system under thermal and pressure controll (NPT, Nose-Hoover
thermostat, prestostat).

 The AIREBO (Adaptive Intermolecular Reactive Empirical Bond Order) 
potential
has been used for the graphene sheet \cite{airebo}.
For the Ru substrate, a recent embedded atomic method (EAM) \cite{EAM:Ru} potential is exploited.
For the C-Ru interaction we developed a new angular dependent Tersoff-like
angular-dependent potential.
In the Tersoff potential file the C-C and Ru-Ru interactions are ignored (nulled out).
The CRuC and RuCRu out-of-plane bond angles were considered only. 
The RuCC and CRuRu angles (with in-plane bonds) are ignored in the applied model.
The consideration of these angles requires the specific
optimization of angular parameters which leads to the polarization
of angles and which does not fit to the original tersoff model.
The details of the fitting procedure can be found in the Supplementary material
of this article.

\begin{figure*}[hbtp]
\label{F1}
\begin{center}
\includegraphics*[height=6cm,width=9cm,angle=0.]{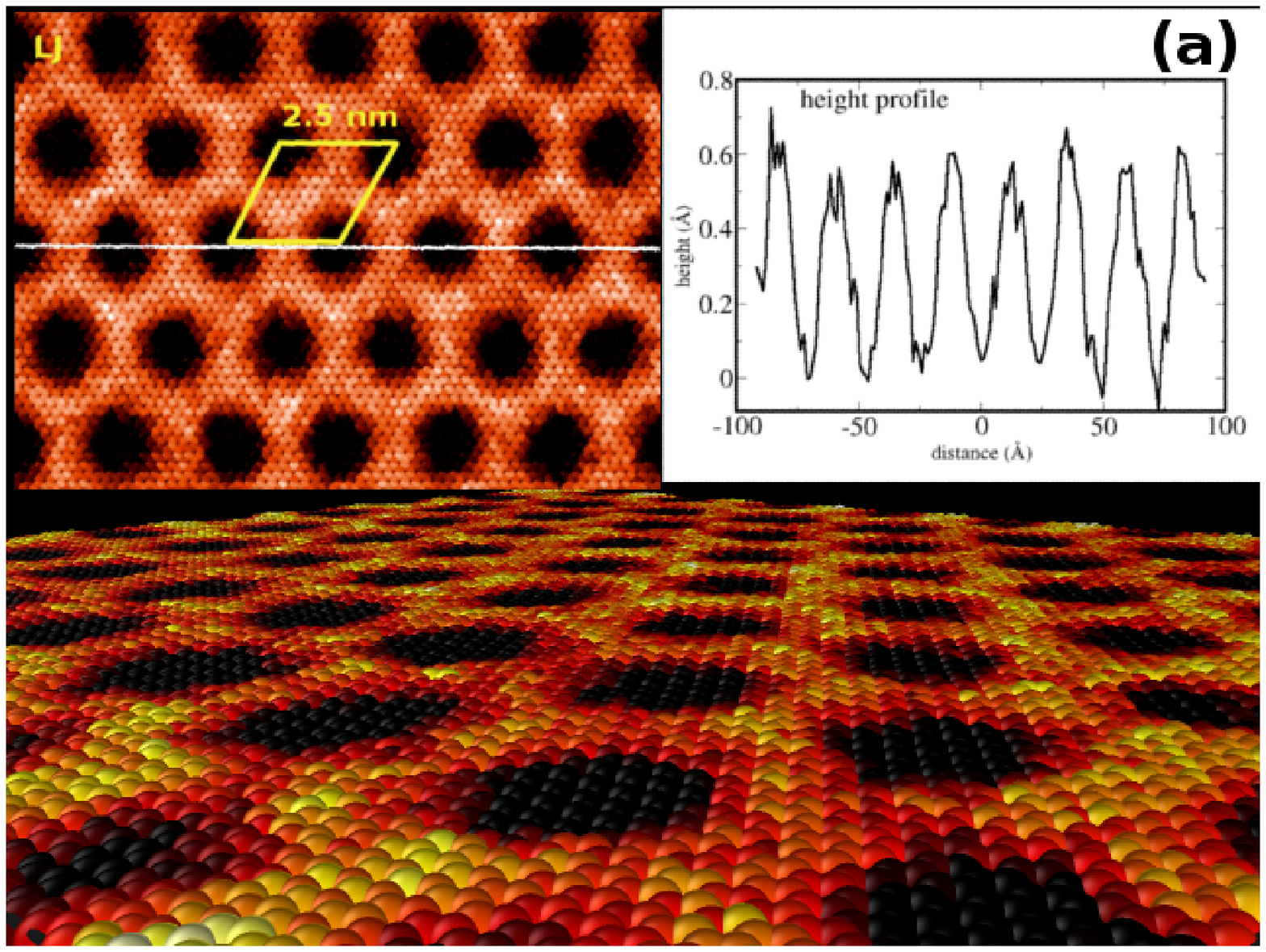}
\includegraphics*[height=6cm,width=8cm,angle=0.]{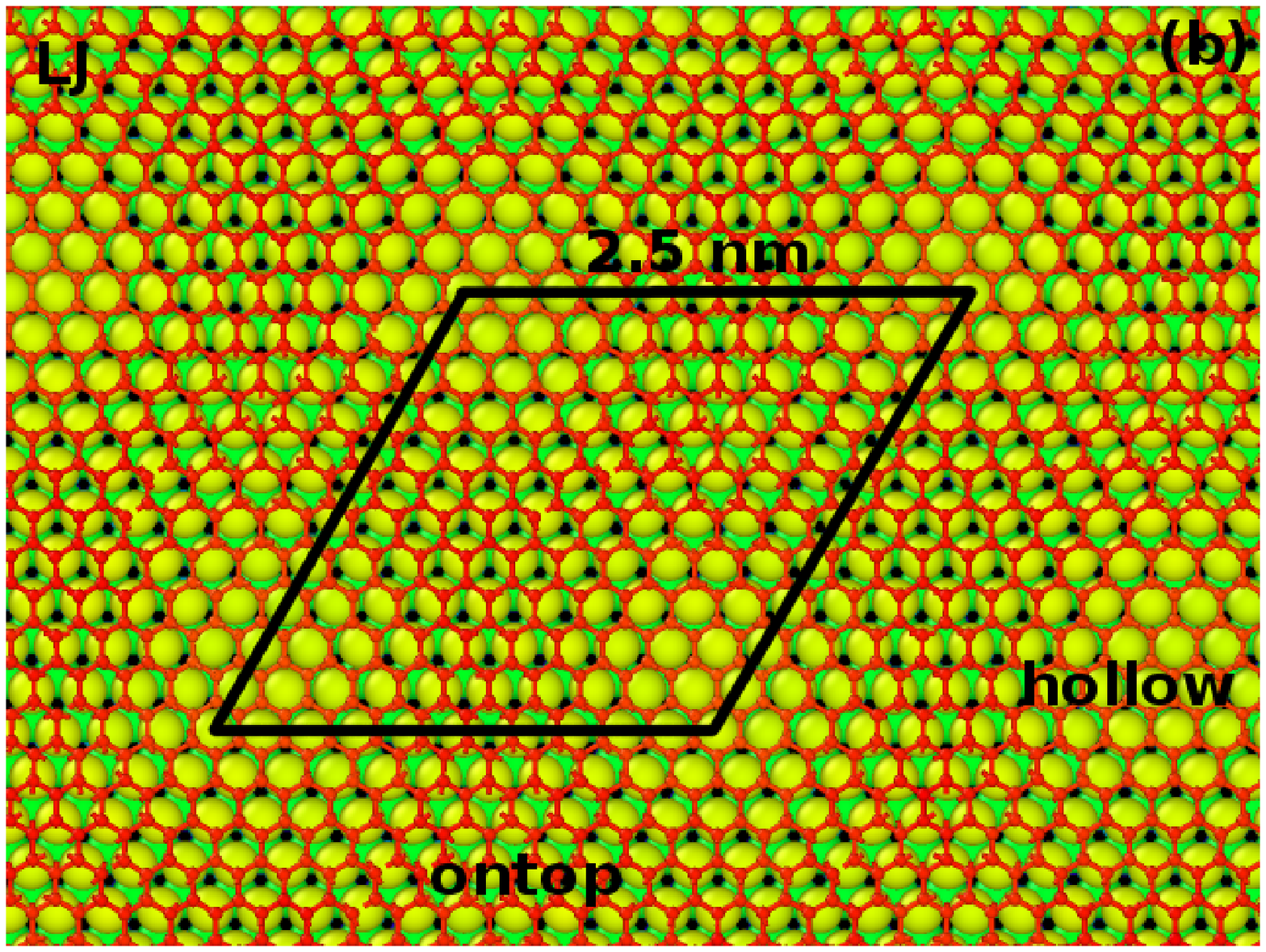}
\caption[]{
 The results of CMD simulations with Lennard-Jones potential
($\epsilon=0.15$, $\sigma=2.1$) at the interface.
(a) The rhomboid supercell is shown ($12 \times 12$ unit cell,  $2.5$ nm).
The top-view with the nanomesh-like topography (black regions are the
bumps).
The landscape view is also shown.
{\em Color coding}: light colors correspond to protrusions (humps) and dark 
ones to bulged-in regions (bumps).
The height profile is also shown along the horizontal line (dissolved white).
The adhesion energy is -0.24 eV/C.
(b) The rhomboid supercell is shown with various registry sites.
Note that the hollow sites correspond to the bumps (valleys)
and the atop positions to the humps (protrusions).
This is an incorrect registry since the reverse is the right one
according to DFT calculations \cite{DFT:Ru-Hutter}.
ones to bulged-in regions (bumps).
}
\end{center}
\end{figure*}


 Before the MD simulations geometry optimization has been applied
using the conjugate gradient algorithm together with the box/relax option
which allows 
the simulation box size and shape to vary during the iterations of the minimizer so that the final configuration will be both an energy minimum for the potential energy of the atoms, and the system pressure tensor will be close to the specified external tensor.

 Isobaric-isothermal (NPT ensemble) simulations (with Nose-Hoover thermostat and a prestostat) were carried out at 300 K, vacuum regions were inserted
between the slab of the gr-substrate system to ensure
the periodic conditions not only in lateral directions (x,y) but also in perpendicular direction to the gr sheet (z).
The variable time step algorithm has been exploited.
Few bottom layers of the substrate have been fixed in order to
rule out the rotation or the translation of the simulation cell.
The code OVITO \cite{ovito}
has been utilized for displaying atomic and nanoscale structures \cite{Sule_2011}.
 The system sizes of $5 \times 5$ nm$^2$ up to $100 \times 100$ nm$^2$
have been simulated under parallel (mpi) environment.

\begin{table}[t]
\raggedright
\caption[]
{
The peak-to-peak corrugation ($\hbox{\AA}$) obtained for gr/Ru(0001)
by various experimental or DFT methods and compared with
CMD results obtained for dome-like superstructures.
}
\begin{ruledtabular}
\begin{tabular}{lcc}
 method (EXP, DFT)  & gr & Ru (topmost) \\
\hline
SXRD-1 & $1.5$ & $0.2$ \\
SXRD-2 & $0.82 \pm 0.15$ & $0.19 \pm 0.02$ \\
LEED & $1.5$ & $0.23$  \\
HAS & $0.15-0.4$ & \\
STM & $0.5-1.1$ & \\
DFT-PBE-1 & $1.75$ & \\
DFT-PBE-2 & $1.6$ & $0.05$ \\
DFT-vdW1 & $1.17$ & \\
DFT-vdW2 & $1.2$ & $0.045$  \\
\hline \hline
present work &   &  \\
\hline
CMD (Morse) & $1.1 \pm 0.1$  & $0.2 \pm 0.05$ \\
CMD (LJ) & $0.5 \pm 0.1$  & $0.3 \pm 0.1$   \\
CMD (Tersoff) & $1.0 \pm 0.2$  & $0.4 \pm 0.2$  \\
T,min (Tersoff)  & $1.0 \pm 0.1$  & $0.35 \pm 0.15$  \\
\end{tabular}
\footnotetext[1]{
SXRD-1: surface X-ray diffraction \cite{SXRD:25X25}, 
SXRD-2: \cite{Xray:Ru}, LEED: Low-energy electron diffraction \cite{LEED:Ru},
HAS: Helium Atom Scattering \cite{Parga}, 
STM: Scanning Tunelling Microscope \cite{STM:Ru}, 
DFT-PBE-1 \cite{DFT:Ru_corrug},
DFT-PBE-2 \cite{DFT:Ru_Wang},
DFT-vdW1 : Density Functional Theory (PBE-Grimme, D2) see in ref. \cite{DFT:Ru-Hutter},
DFT-vdW2 (PBE-Grimme, D2) see in ref. \cite{DFT:Ru-Stradi},
CMD: classical molecular dynamics (present work) obtained
using the Morse and LJ potentials.
{\em T,min}: geometry optimization and energy minimization only (molecular mechanics, no MD) using the steepest-distant
method for finding the energy minimum of the system together
with anisotropic box relaxation (Tersoff-only). This method provides results
which can be compared directly with {\em ab initio} DFT geometry optimization
results.
DFT/vdW-DF2/CMD: The DFT/revPBE-DF2 vdW-functional \cite{revPBE,Dion} is used
as imlemented in the trunk version of SIESTA (LMKLL) \cite{SIESTA2}
including 748 atoms (302 Carbon atoms) with a minimal supercell.
}
\label{T1}
\end{ruledtabular}
\end{table}

\subsection{{\em Ab initio} DFT calculations}

 First principles DFT calculations have also been carried out for
calculating the adhesion energy per Carbon atoms
vs. the C/Ru distance for a small ideal system with a flat graphene layer.
The obtained potential energy curves
can be compared with the similar curve of MD calculations.

 For this purpose we used 
the SIESTA code \cite{SIESTA,SIESTA2} which utilizes atomic centered numerical basis set.
The SIESTA code and the implemented Van der Waals functional (denoted as DF2,
LMKLL in the code \cite{SIESTA2} 
have been tested in many articles for gr (see e.g recent refs. \cite{sies1,sies2}).
We have
used Troullier Martin, norm conserving, relativistic pseudopotentials in fully separable
Kleinman and Bylander form for both carbon and Ru.
Double-$\zeta$ polarization (DZP) basis set was
used.
In particular, 16 valence electrons are considered for Ru atoms
and 4 for C atoms.
Only $\Gamma$ point is used for the k-point grid in the SCF cycle.
The real space grid used to calculate the Hartree, exchange and correlation
contribution to the total energy and Hamiltonian was 300 Ry (Meshcutoff).
 The gradient-corrected Exchange and correlation are calculated
by the revPBE/DF2 van der Waals functional \cite{Dion}.
The Grimme's semiempirical functional \cite{Grimme} has also been used together
with the PBE/GGA DFT functional \cite{PBE}.
 The system consists of 299 Carbon and 233  Ru atoms (3 layers Ru).  

\section{Results and Discussion}

\subsection{Results for pairwise potentials}

 First results will be shown briefly obtained by the simple
Lennard-Jones potential (see Fig. 1(a)-(b)).
The LJ potential not only provides wrong registry but also
the supercell size is too small: $\sim 2.5$ nm.
The shape of the rhomboid supercells are somewhat distorted and the topography becomes
disordered at lower adhesion energy ($E_{adh} < -0.25$ eV/C).
The corrugation is too low ($\xi \approx 0.08$ nm).

 Not only LJ, but other simple pair-wise potentials, such as Morse
 potential
is unable to reproduce Moir$\acute{e}$ superstructures.
After visual inspection of Figs. 1 one can conclude that
the LJ potential provides also incorrect topography: it favors energetically
hollow site vs. on-top configurations (hollow bumps and ontop humps).
This is again in contrast with DFT results which show the contrary results (hollow humps and ontop bumps) \cite{Batzill}.
Physical intuition also suggests the stronger adhesion of ontop positions
where 3 Carbon atoms within a Carbon hexagon are in close contact with 3 first neighbor Ru atoms and the rest of the Carbon atoms bind only to 2nd
neighbor Ru atoms \cite{Batzill}. 
In the hollow configuration a first neighbor Ru atom is in the
middle of the hexagon and 3 2nd neighbor (2nd layer) Ru atoms
are nearly covered by Carbons \cite{Batzill}. The hollow registry naturally
lead then to weaker adhesion. 
Contrary to this,
CMD-Morse and LJ overbind hollow sites vs. DFT which favors fcc and hcp atop sites.
CMD gives the adsorption energy of -0.222 eV/C for a purely hollow registry, -0.082 eV/C for atop configurations and -0.172 eV/C for a mixed system (hollow and other sites).
The Morse function gives, though the
supercell size is correctly as $2.7$ nm.
 The peak-to-peak corrugation ($\xi_{corr}$) is averaged for small subregions which
includes the $12 \times 12$ superstructure.
The sampling of these regions was taken on large systems with a lateral size
up to $100 \times 100$ nm$^2$.
We find that the overall average surface corrugation remains 
in the range of $\xi \approx 1.1 \pm 0.1$ $\hbox{\AA}$ 
when the adhesion energy of the gr-layer is
$E_{adh} \approx -0.17$ eV/C atom.

  The corresponding literature data on corrugation obtained by various experimental and DFT methods
are summarized in Table ~\ref{T1}.
In spite of the great efforts to obtain the structural corrugation
of the moir$\acute{e}$ observed in gr/Ru, the magnitude of
the corrugation is still a subject of controversy.
In general, the data
scatters within the wide range of $0.15-1.75$ $\hbox{\AA}$.
The lower range is provided mostly by
HAS \cite{Parga} and SXRD \cite{SXRD:25X25} experiments and partly by STM \cite{STM:Ru}.
Other experiments, such as SXRD \cite{Xray:Ru}, STM \cite{STM:Ru}, LEEM \cite{LEED:Ru}  and DFT methods \cite{DFT:Ru_corrug,DFT:Ru_Wang},
provided larger corrugation values in the range of $1.2-1.75$ $\hbox{\AA}$.
  STM values are strongly influenced by the applied bias voltage \cite{Parga,STM:Ru}
providing corrugation data in the range of $0.5-1.1$ $\hbox{\AA}$.

\begin{figure*}[hbtp]
\begin{center}
\includegraphics*[height=6cm,width=8cm,angle=0.]{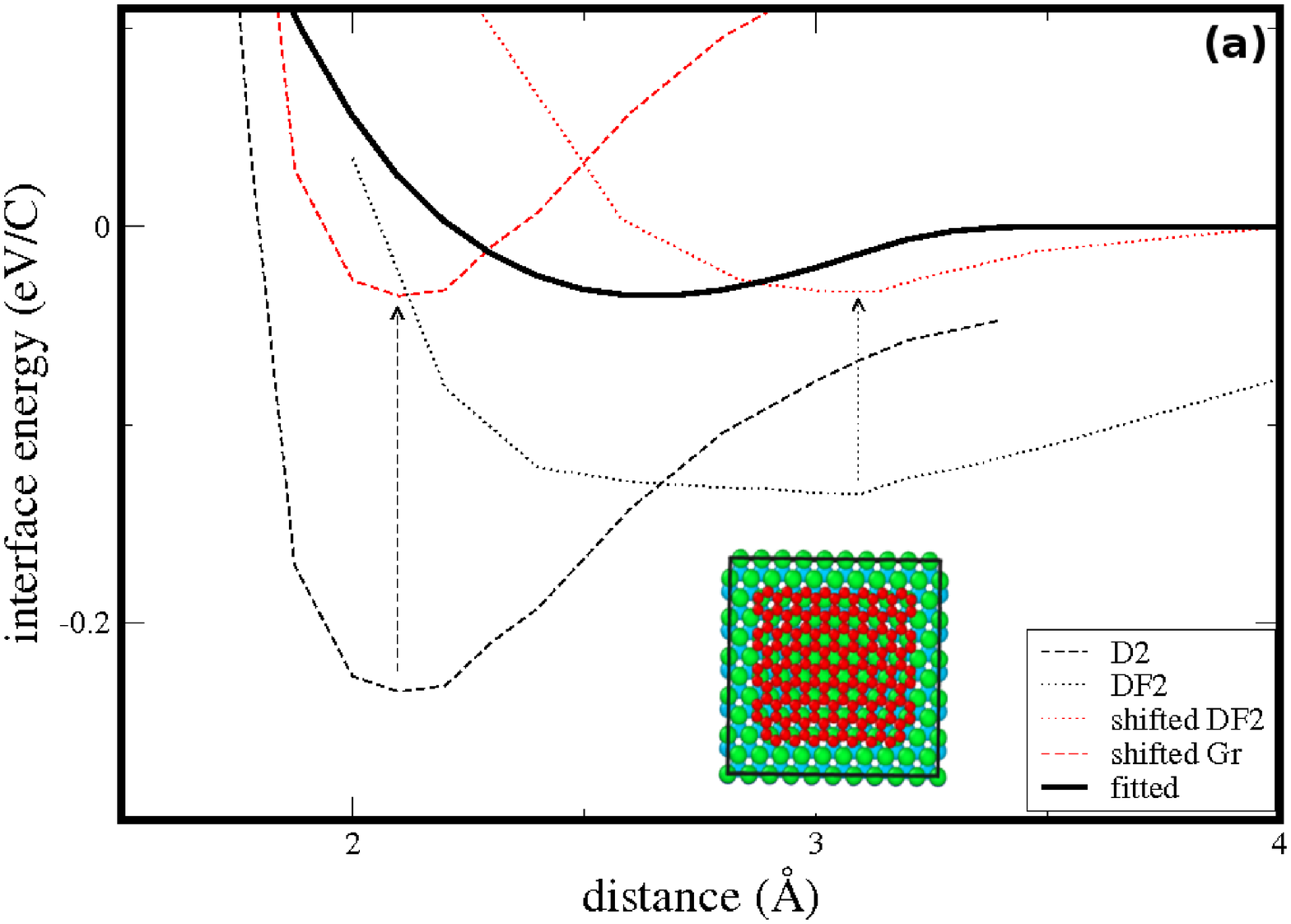}
\includegraphics*[height=6cm,width=8cm,angle=0.]{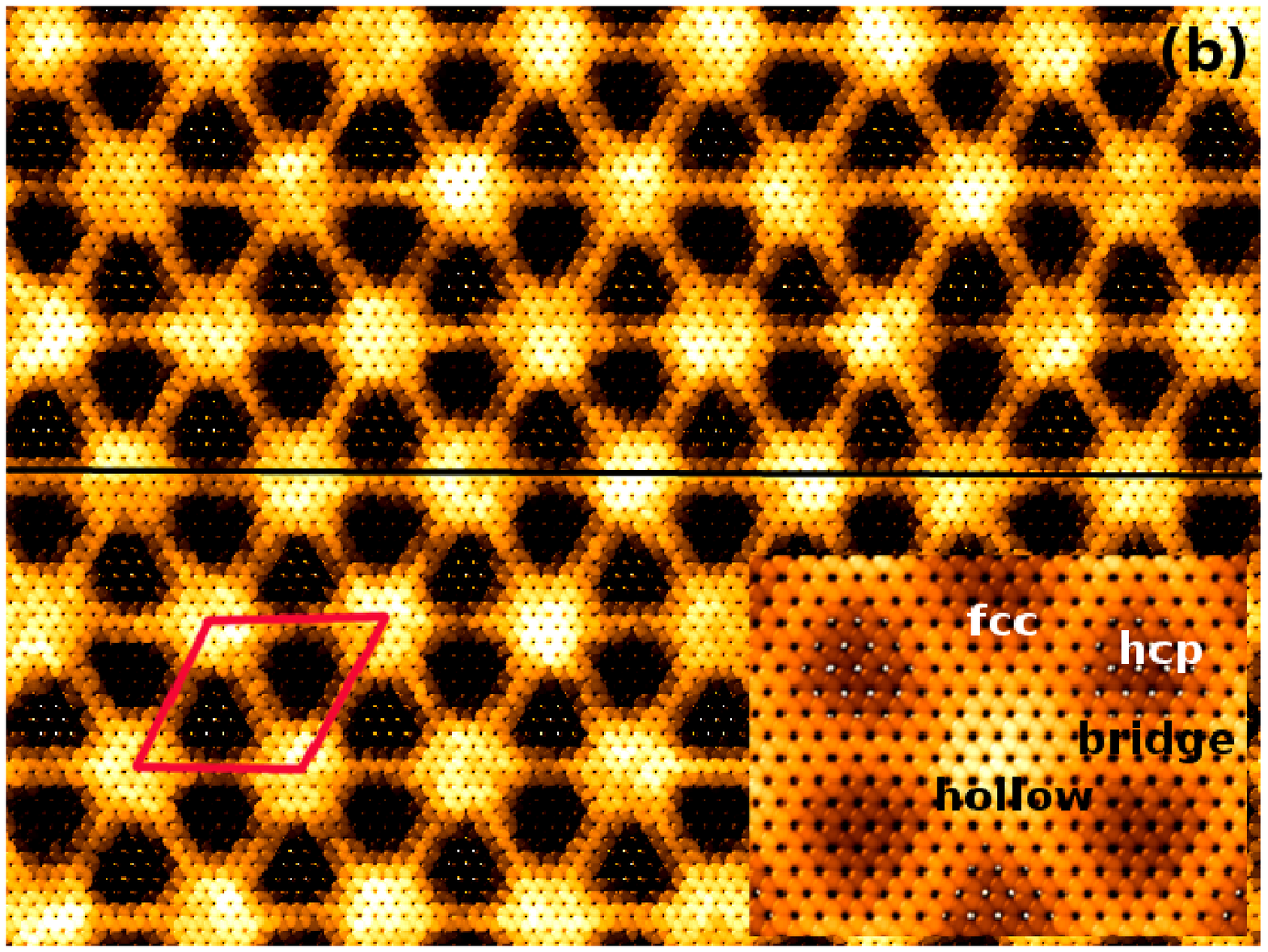}
\includegraphics*[height=6cm,width=8cm,angle=0.]{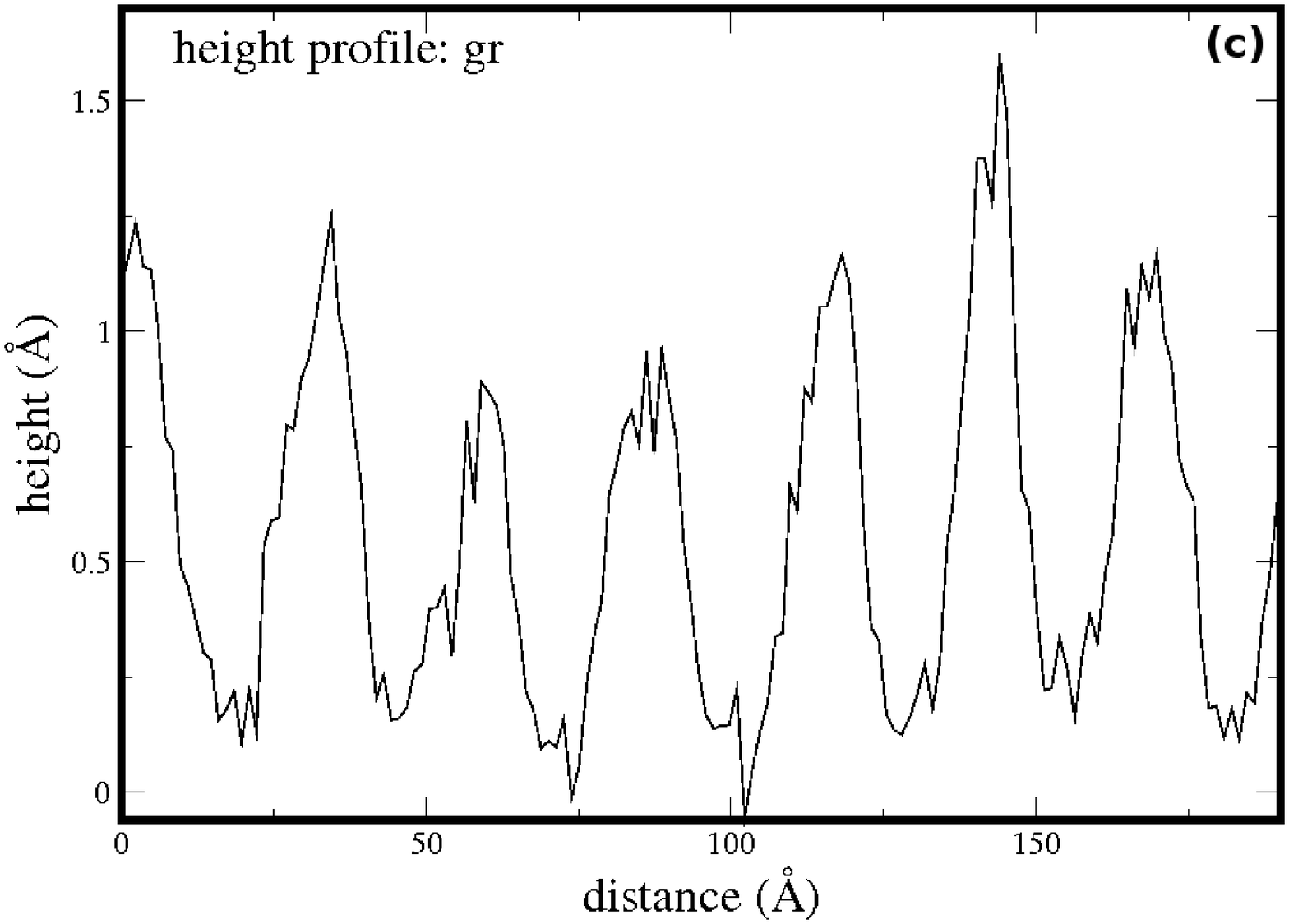}
\includegraphics*[height=6cm,width=8cm,angle=0.]{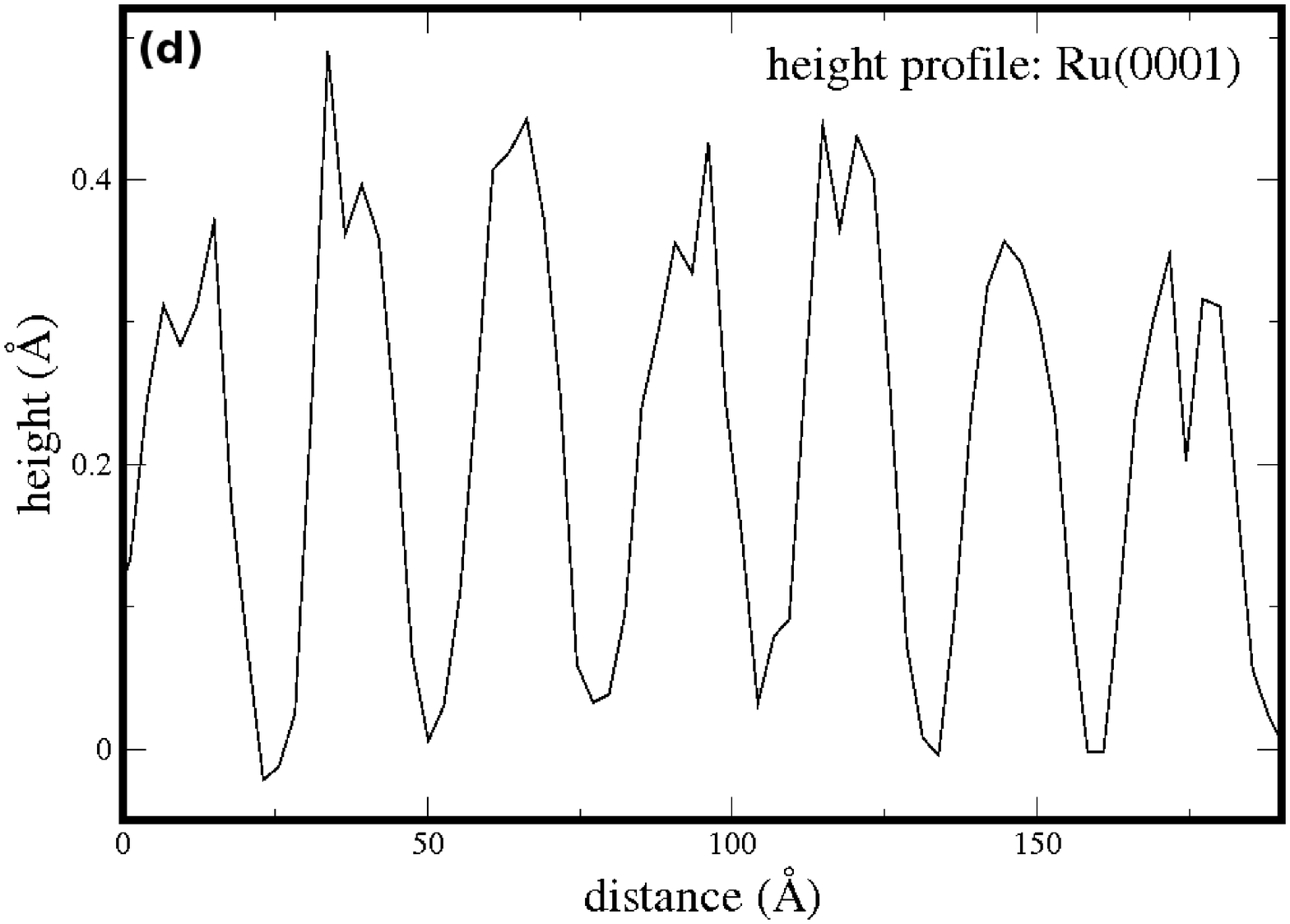}
\caption[]{
(a) The potential energy curves (eV/Carbon)
as obtained for a small flat graphene/Ru(0001) system (562 atoms) supported on
a rigid Ru substrate.
The distance denotes the interfacial separation (adsorption distance, $\hbox{\AA}$) of the
flat gr-layer from the topmost layer of Ru(0001).
DFT-D2 and DF2 correspond to {\em ab initio}
DFT calculations using the Grimme's semiempirical dispersion
function (with standard parameters) \cite{Grimme}
and the non-local vdw functional (LMKLL, \cite{LMKLL}) as implemented in the code
SIESTA \cite{SIESTA}.
The shifted DFT potential energy curves are also shown (dashed and dotted curves).
(b) The obtained moire superstructures after precise conjugate gradient geometry optimization using the new fitted force field.
In the inset a zoomed region is shown with a hollow hump in the middle.
The various registries are also shown (fcc, hcp and bridge).
Color coding is similar to that used in Fig. (1).:
(c)-(d) Height profiles for the graphene and the Ru(0001) surfaces, respectively,
cut along the x-axis in the middle of Fig 2(b).
}
\end{center}
\label{F2}
\end{figure*}

\begin{table*}[tb]
\raggedright
\caption[]
{
The summary of various properties obtained for gr/Ru(0001)
by classical molecular dynamics simulations using the fitted Tersoff
potential for the interface.
The properties of moire superstructures.
}
\begin{ruledtabular}
\begin{tabular}{lccccccccc}
  &  &  &  &  &  &  &  &  &   \\
  pw & $\xi$ ($\hbox{\AA}$)  & $\xi_{Ru}$ ($\hbox{\AA}$) & $d_{min}$ ($\hbox{\AA}$) & $d_{max}$ ($\hbox{\AA}$) & $a_{gr}$ ($\hbox{\AA}$) & $a_{lm}$ ($\%$) & $E_{adh}$ (eV/C) & $E_{str}$ (eV/C) & $E_{gr}$ (eV/C) \\
  &  &  &  &  &  &  &  &  &   \\
\hline
  &  &  &  &  &  &  &  &  &   \\
 CMD  & $1.0 \pm 0.2$ & $0.4 \pm 0.2$ & 2.2-2.5 & 2.8-3.1 & 2.47 & 8.70 & -0.17 & 0.0838 & -7.2876 \\
  &  &  &  &  &  &  &  &  &   \\
  &  &  &  &  &  &  &  &  &   \\
 EXP & $0.8 \pm 0.3$ & $0.2 \pm 0.15$ & 2.0-2.1 & 2.8,3.6 & 2.45 & 10 & n/a & n/a & n/a \\
  &  &  &  &  &  &  &  &  &   \\
  &  &  &  &  &  &  &  &  &   \\
 DFT & 1.1-1.7$^b$ & 0.05$^b$ & 2.0$^{b,c}$ & 3.8$^{b,c}$ & 2.51-2.54$^c$ & 
 6-7$^c$ & -0.20$^{b,c}$ & 0.104$^{b,c}$ & n/a   \\
  &  &  &  &  &  &  &  &  &   \\
\end{tabular}
\footnotetext*[1]{
pw denotes present work,
$\xi$ and $\xi_{Ru}$ are the average corrugation for gr and the topmost
Ru(0001) layer ($\hbox{\AA}$).
$d_{min}$ and $d_{max}$ are the minimal and maximal inter-layer distances ($\hbox{\AA}$) at the interface.
$a_{gr}$, $a_{lm}$ are the lattice constant of gr ($\hbox{\AA}$) and
the lattice mismatch ($\%$) after simulations ($a_{lm}=100 (a_{s}-a_{gr})/a_{gr}$).
\\
CMD: Tersoff-dispersion results (Tersoff-Grimme potential)) with CMD at 300 K.
\\
EXP: the cumulative experimental results are also given which are obtained
on the basis of data shown in Table I.
corrugation ($\xi$): SXRD and STM results \cite{Xray:Ru,STM:Ru}, 
$d_{min}$: from refs. \cite{Xray:Ru,LEED:Ru},
For $d_{max}$ the lower and higher limit are obtained from refs.  
\cite{Xray:Ru} and \cite{LEED:Ru}, respectively.
\\
DFT results are also given for comparison \cite{DFT:Ru-Hutter,Batzill,DFT:Ru_Wang}.
DFT (pw): PBE \cite{PBE} and revPBE/vdW-DF2 \cite{Dion} results as obtained by the author. See further details
in the caption of Table I.
All quantities are given per Carbon atom, except the
atomic cohesive energy of the substrate.
\\
The adhesion energy $E_{adh}=E_{tot}-E_{no12}$, where $E_{tot}$ is the potential energy/C
after md simulation. $E_{no12}$ can be calculated using the final
geometry of md simulation with heteronuclear interactions
switched off. Therefore, $E_{adh}$ contains only contributions from
interfacial interactions.
$E_{str,gr}$, the strain energy of the corrugated gr-sheet
include terms comming from stretching and corrugation (bending and torsional strain)
($E_{str,gr}=E_{gr}-E_{gr,flat}$),
where $E_{gr}$ and $E_{gr,flat}=-7.3715$ eV/C are the cohesive energy of C atoms
in the corrugated and in the relaxed flat (reference) gr sheet as obtained
by the AIREBO C-potential \cite{airebo}.
\\
$^b$ from ref. \cite{DFT:Ru-Hutter},
$^c$ from ref. \cite{DFT:Ru_corrug}.
}
\label{T2}
\end{ruledtabular}
\end{table*}

\begin{figure*}[hbtp]
\begin{center}
\includegraphics*[height=6cm,width=8cm,angle=0.]{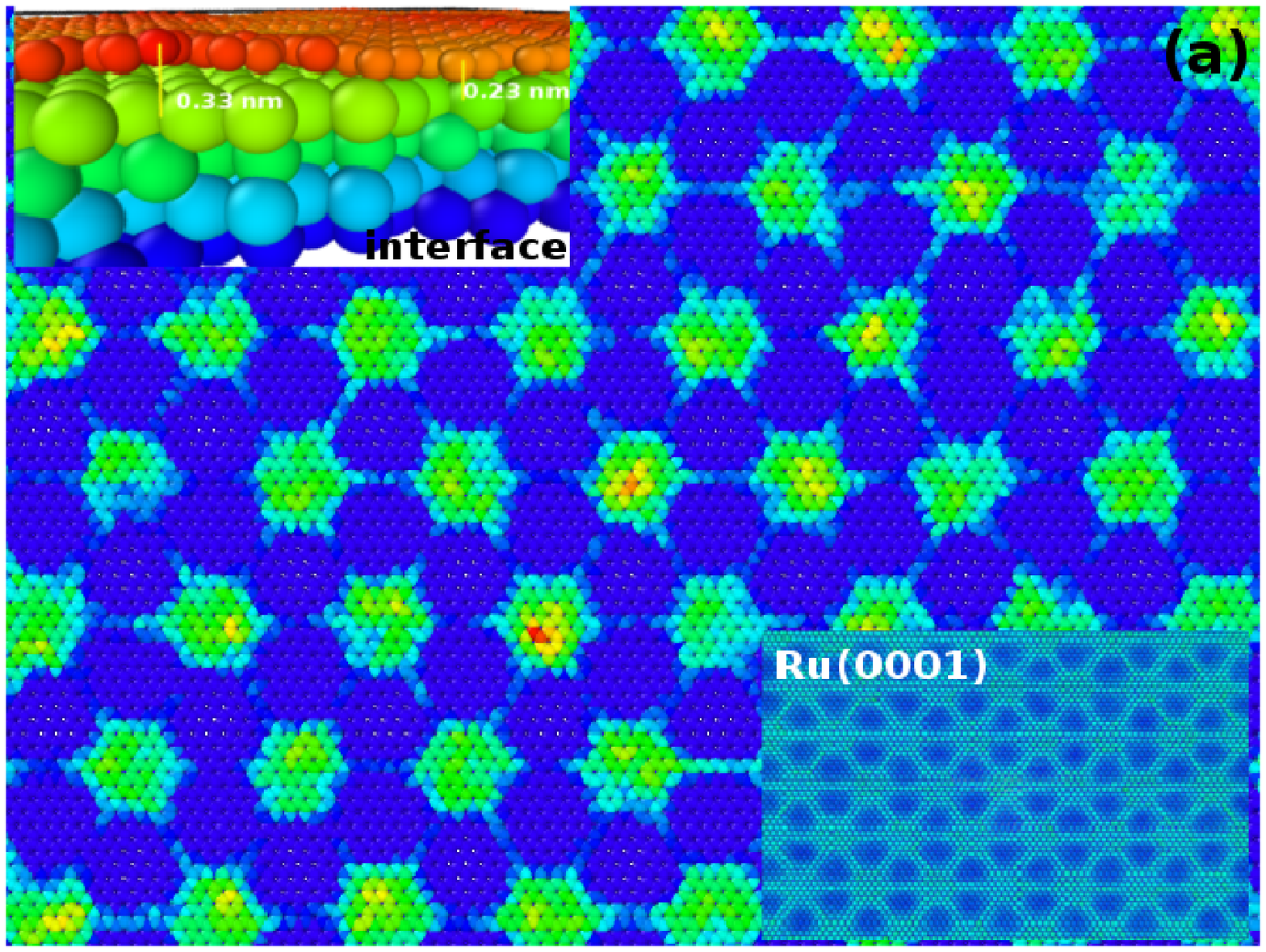}
\includegraphics*[height=6cm,width=8cm]{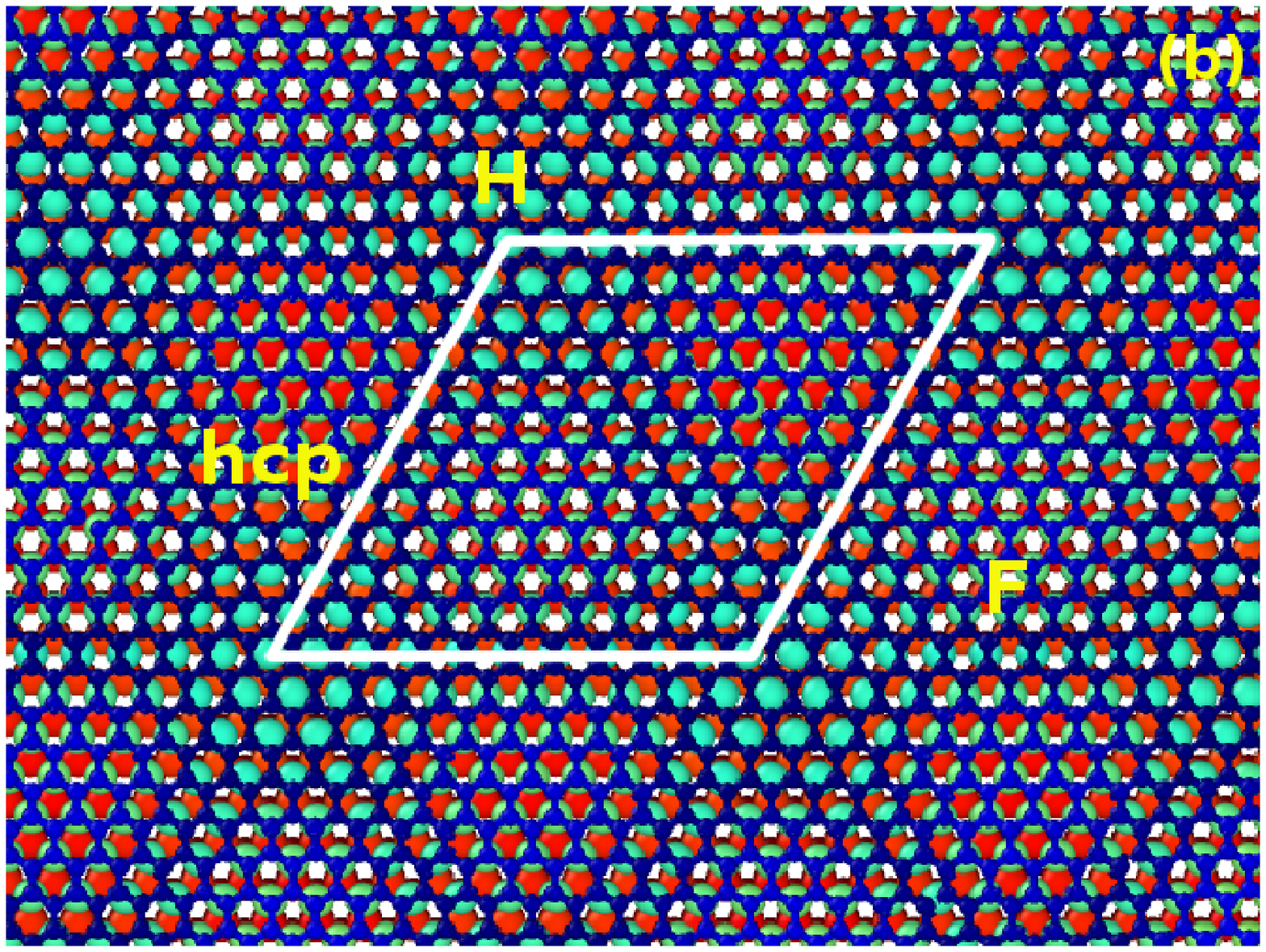}
\caption[]{
A large-scale gr/Ru(0001) system as obtained by classical molecular dynamics simulations
using the new Tersoff interfacial potential at 300 K.
(a) top view with color coding which helps to characterize
the hum-and-bump superstructure.
Lower inset: the surface of the topmost layer Ru(0001).
Upper inset: the interface cut in the middle of the simulation cell.
Color coding is similar to that used in Figs. (1)-(2).
(b) A wireframe modell of the $12 \times 11$ supercell with binding registries.
H: hollow sites, F: fcc ontop sites, hcp: hcp ontop sites
}
\end{center}
\label{F5}
\end{figure*}

\subsection{Results for the Tersoff interface}


 The employed new C-Ru force field has been parameterized for
fictitious C-Ru alloy systems. The training set does not contain
information directly on the corrugation of graphene on Ru(0001).
Hence the obtained morphology of gr can be taken as an
independent property of the developed force field (quasi "a priori" result).
In other words we did not force by hand via the employed parameters the interfacial force field
to reproduce the experimental corrugation and morphology.
This is remarkable because simple imposed conditions were sufficient
to develop an adequate gr/Ru(0001) interfacial potential.

 The details of parameter fitting (PF) is given in the Supplementary Material.
Hereby we outline only the 3rd step of parameter fitting where we
considered ab initio DFT potential enegy curves (PECs) of
small modell gr/Ru(0001) systems.
In the 3rd step of PF the force field has been forced to
fit to potential enegy curves (PECs) obtained by DFT calculations.
On Fig. 2(a) the
PECs can be seen as calculated for
a small modell system including 520 atoms.
The gr-Ru(0001) inter-layer distance has been varied
systematically in order to scan the potential energy surface
at the interface.
We kept gr completely flat, hence no corrugation is taken into
account in this simple modell as well.
It has been found that various {\em ab initio}
DFT curves differ from each other considerably.
The widely used DFT-D2 functional \cite{Grimme} gives an
energy minimum at $d_{CRu} \approx 2.1$ $\hbox{\AA}$ (similarily to LDA PEC \cite{RPA}) while the nonlocal DF2 (LMKLL \cite{LMKLL})
at $d_{CRu} \approx 3.0$ $\hbox{\AA}$.
It is known for a while that the nonlocal DF and DF2 tends be too repulsive at short distances
which leads to overestimated equilibrium distances \cite{RPA}.
On the other hand the semiempirical Grimme's function provides 
overbinding leading to too short equilibrium distances and
to overestimated adhesion energy \cite{RPA}.
Recent accurate DFT calculations using the Random Phase Approximation
(RPA) for correlation predict in many cases that the RPA potential energy curves
are between the curves of LDA and vdw-DF2 functionals \cite{RPA}.
In our case the DFT-D2 seems to overbind gr-Ru(0001) in a similar way as
LDA does. vdW-DF2, however, puts the equilibrium distance too far ($r_0 \approx 3.1$ $\hbox{\AA}$). Unfortunately, we found
no data in the literature for gr/Ru(0001) with RPA.
For a very similar system (gr/Ni(111)) RPA gives a PEC
between LDA and DF2.
The nearly correct PEC can also be expected somewhere
between the Grimme and DF2 curves for gr/Ru(0001) since this
system behaves in a very similar way to gr/Ni(111), e.g. the adhesion energy is in the
same range.

 After much effort we find that the original DFT PECs are insufficient
for parameter fitting.
 The nonlinear least square fit either to Grimme's-D2 or vdW-DF2 curves give model force fields
which lead to a wrong nanomesh-like morphology similar to that of pairwise potentials and to overbinding (the adhesion energy becomes too large).
We extensively tested this situation using a code written for this purpose \cite{potfit},
however attempts were failed to reach a satisfactory force field.
Either if the tersoff potential has been fitted to equilibrium
DFT geometry or to DFT PECs the adhesion energy becomes too negative
even if the topography is already nearly correct.
The divergence of various standard DFT potential energy curves lead us to
follow an another strategy at parameter fitting (RPA goes beyond
the scope of the present paper since it is extremely time consuming).
Fitting to artificial weakly bound alloys (step 2 in fitting with the code PONTIFIX \cite{pontifix}, details are given in
the Supplementary Material) instead of DFT PEC, provided, however, an adequate
surface pattern.
Using this way of parameter fitting
we got a PEC 
which is positioned somewhere between the two DFT curves with an energy
minimum at $d_{CRu} \approx 2.5$ $\hbox{\AA}$ (2nd step of parameterization).

 As a further refinement step, we also optimized the parameter set with a fitting
code written for this purpose in our Lab \cite{potfit} using the modified DFT PECs (details see in the
Supplementary Material).
It is remarkable, that considering an additional reasonable training
set (small modell systems with flat gr and corresponding modified DFT PECs)
the pre-optimized force field
(as obtained by the code PONTIFIX)
can be further refined (mostly the morphology and the magnitude of adhesion).
However, the DFT PECs had to be shifted upwards (see Fig 2(a)) in order
to get a reasonable training set for fitting and to avoid
overbinding.
Both the D2 and DF2 functionals seem to overestimate adhesion at completely
different equilibrium distances.
Therefore, the PECs are shifted upwards by $\sim 0.2$ and $\sim 0.1$ eV/C,
respectively.
The upward shift of the PECs is further rationalized by the fact
that our cutoffed short-range tersoff potential accounts for only
first neighbor interactions which seriously reduces
the depth of the potential energy curve.
In this way we are able to tune the adhesion energy which is hard to be
achieved in step 2.

 In this case we tried to force the preoptimized force field 
to be adapted somewhere between the scaled-up (shifted) DFT PECs where
we expect the corresponding more accurate RPA curve \cite{RPA}.
It has been found that such fitting conditions keep
the essential properties of the previously optimized force field
(adhesion energy, hump-and-bump morphology with hollow humps).
We find that the flat gr/Ru(0001) system is more repulsive (weakly
adhesive, $E^{flat,gr}_{adh} \approx -0.035$ eV/C) than provided by D2 and DF2 DFT functionals in accordance with 
results obtained for other systems \cite{RPA}.
The measured stronger adhesion ($E^{corr,gr}_{adh} \approx -0.2$ eV/C) 
could only appear in the corrugated system.

 In general, one can say that using a fitted Tersoff function
for the gr-metal couple, the overall landscape and topography
of the moir$\acute{e}$ superstructure changes significantly.
The bump ruled pattern (looks like a nanomesh) of the pure Morse or LJ potential dissapear and the correct "hump-and-bump" like features
 with {\em hollow-humps} (protrusions) and ontop-bumps (bulged-in regions) can be seen. Hence, using bond order potentials
with angular dependence at the interface
the topography becomes essentially correct which is already comparable
with experimental STM results \cite{STM:Ru}.
The most important properties of the supercells, the size and corrugation
are perfectly in accordance with experiment (supercell size: 2.7 nm)
or in reasonable agreement with many of the experimental and DFT methods.
In particular, the corrugation obtained by the Tersoff-only function
is in the range of $\xi \approx 1.1$ $\hbox{\AA}$ at 300 K which is
somewhat smaller than the average value obtained by vdw-DFT methods
($1.2$ $\hbox{\AA}$).
Moreover, $\xi$ is very close to
surface X-ray diffraction measurements \cite{Xray:Ru}
and to most of the STM results \cite{STM:Ru} (see overiew of results
in Table I.).

 In Table II. the interfacial atomic distances are also given in which gr-topmost Ru(0001)
distances ($d_{min}$ and $d_{max}$) are shown. Within the humps (flat domes) 
$d_{max} \approx 2.9 \pm 0.2$ $\hbox{\AA}$ while at the bumps $d_{min} \approx 2.3 \pm 0.2$ $\hbox{\AA}$.
This is similar to that found by DFT ($3.3$ and $2.2$).
The corrugation of the topmost Ru-layer is somewhat larger than
provided by DFT \cite{DFT:Ru-Stradi}  or SXRD \cite{Xray:Ru} ($\xi \approx
0.4 \pm 0.1$ $\hbox{\AA}$).
It could be that the employed Ru EAM potential \cite{EAM:Ru} 
is not rigid enough at the (0001) surface.
However, the only known deficiency of this potential
is the underestimation of the melting point, other properties, such
as surface energy, which is important in this case, is nicely reproduced
\cite{EAM:Ru}.
The simulated lattice missfit is at around 8.7 $\%$ (DFT values: 6-7 $\%$)
which is due to the significant surface reconstruction
of the topmost Ru(0001) layer (see inset Fig. 3(a)).
The system is less distorted than the
experimental missfit of 10 $\%$
based on idealistic lattice constants
in the separated materials ($a_{gr} = 2.46$ $\hbox{\AA}$,
and $a_{Ru}=2.706$ $\hbox{\AA}$).
The lattice constant in the topmost Ru layer is 2.69 $\hbox{\AA}$ vs.
the bulk value of 2.706 $\hbox{\AA}$.

 The strain energy of the gr-sheet (the energy difference between the corrugated and flat gr) has also been calculated and found to
be smaller (0.02 eV/C) than that of obtained by DFT calculations (0.1 eV/C).
This could be due to the smaller corrugation and to the
smaller energy difference between the flat and corrugated gr-sheet.

\section{Conclusions}

 Using classical molecular dynamics simulations we have shown
that it is possible to provide reasonably accurate results for weakly bound
extended systems such as gr/Ru(0001).
To achieve this task, however, it was important to develop
a new force field which adequately describe weak bonding between
the gr-sheet and Ru(0001).
It turned out that the widely used simple pair potentials (such as Lennard-Jones 
and Morse) lead to improper C-Ru bonding orientations and favor
incorrectly hollow-site registries instead of on-top ones.
The overall topography becomes then wrong: bumps (bulged-in regimes) can be found
at hollow-sites and the humps (bulges) at on-top registries (fcc and hcp positions).

 Properly oriented adhesion could only be reached with
a newly parameterized angular-dependent Tersoff potential for C-Ru interactions.
The obtained adhesion energy is very close to the DFT results.
The simulations automatically lead to the
$(12 \times 12)C/(11 \times 11)Ru$ nearly commensurate superstructures
and the obtained supercell size is in accordance with the available
experimental and DFT results. 
 The calculated corrugation ($\xi \approx 1.1$ $\hbox{\AA}$) is 
in agreement with most of the relevant experimental measurements and
DFT results.
This could be taken as an important result because the obtained
morphology and corrugation is independent from the
employed training data set and hence can not be considered as
a manually loaded property.

 The application of the newly developed force-field could help to explain
the obtained STM micrographs. Moreover the Tersoff gr/Ru(0001) system
can also be used for various computer experiments such as
the simulation of the superior thermal conductivity of gr nanoribbons
or the nanomechanics of supported gr under external load or
for the ion-patterning of graphene.
In general, the new model opens the way towards large-scale supported gr-simulations
under various conditions
which was possible until recently only by simple pairwise potentials
which, however, provide inadequate moire structures as pointed
out in this article.

%
%

\section{acknowledgement}
The calculations (simulations) have been
done mostly on the supercomputers
of the NIIF center (Hungary).
The kind help of P. Erhart (Darmstadt) in the usage of the PONTIFIX code is greatly acknowledged.
The availability of codes LAMMPS (S. Plimpton) and OVITO (A. Stukowski) are also greatly acknowledged.

\appendix
Supplementary Material is available on parameter fitting.


\end{document}


\newcommand{\bec}{\begin{center}}
\newcommand{\ec}{\end{center}}
\newcommand{\be}{\begin{equation}}
\newcommand{\ee}{\end{equation}}
\newcommand{\beqn}{\begin{eqnarray}}
\newcommand{\eeqn}{\end{eqnarray}}
\newcommand{\bet}{\begin{table}}
\newcommand{\ent}{\end{table}}
\newcommand{\bib}{\bibitem}

\newcommand{\sect}[1]{Sect.~\ref{#1}}
\newcommand{\fig}[1]{Fig.~\ref{#1}}
\newcommand{\Eq}[1]{Eq.~(\ref{#1})}
\newcommand{\eq}[1]{(\ref{#1})}
\newcommand{\tab}[1]{Table~\ref{#1}}

\renewcommand{\vec}[1]{\ensuremath\boldsymbol{#1}}
\renewcommand{\epsilon}[0]{\varepsilon}

\newcommand{\cmt}[1]{\emph{\color{red}#1}%
  \marginpar{{\color{red}\bfseries $!!$}}}



\title{
{\bf \LARGE Supplementary Material} \\
\vspace{1cm}
{\em The DFT and molecular dynamics multiscale study of the corrugation of graphene on Ru(0001):
the unexpected stability of the moire-buckled structure
\\
}
\vspace{1cm}
The details of parameter fitting for the gr-Ru(0001) interface
}


\author{P. S\"ule, M. Szendr\H o} \email{sule@mfa.kfki.hu} 

\affiliation
{Research Centre for Natural Sciences,
Institute for Technical Physics and Materials Science
\\
Konkoly Thege u. 29-33, Budapest, Hungary,sule@mfa.kfki.hu,
The University of E\"otv\"os L\'or\'and, Department of
Materials Physics, Budapest.
\\
}

\date{\today}


\begin{abstract}
In this Supplementary Material the details of the
parameter fitting procedure are shown.
In particular, we discuss the least square fitting of the
interfacial C-Ru Tersoff potential which is suitable for
simulating the gr/Ru(0001) weakly bound complex.
\end{abstract}

\maketitle

\section{The Tersoff potential}

 The Tersoff potential is given in the following form:
\be
V_{Tersoff}=\sum_{ij,i>j} f_{ij}(r_{ij})[V_{ij}^R-b_{ij}(\Theta) V^A_{ij}(r_{ij})].
\ee
The radial part of the Tersoff potential is composed of the
following repulsive and attractive functions,
\be
V_{ij}^R=A exp(-\lambda_1 (r-r_0)),
\ee
\be
V_{ij}^A=B exp(-\lambda_2 (r-r_0)).
\ee
The angular dependence is introduced via the attractive part $V_{ij}^A$ term by
the $b_{ij}(\Theta)$.
\be
b_{ij}(\Theta)=(1+\beta^n \chi_{ij}^n(\Theta))^{\frac{1}{2n}}
\ee

\be
\chi(\Theta)=\sum_{k(\neq i,j)} f_{ik}^c(r_{ik}) g_{ik}(\Theta_{ijk})  exp[2\mu_{ik}(r_{ij}-r_{ik})]
\ee
where the angular term $g(\Theta)$,
\be
g(\Theta)=\gamma \biggm(1+\frac{c^2}{d^2}-\frac{c^2}{d^2+(cos\Theta-h)^2}\biggm),
\ee
where $h=cos(\Theta_0)$. We find that $\Theta_0 \approx 80^{\circ}$ is the
most favorable C-C...Ru bond angle for graphene.
The cutoff function is
\[
f_{ij}(r_{ij})= \left\{ \begin{array}{cc}
~~~~~~~~~ 1 & r \le R_c-D_c  \\ \frac{1}{2}-\frac{1}{2}sin[\frac{\pi}{2}(r-R_c)/D_c] & |r-R_c| \le D_c
 \\
0 & r \ge R_c+D_c
\end{array} \right. \]
where $R_c$ $\hbox{\AA}$ is the cutoff distance
and $D_c$ $\hbox{\AA}$ is the damping distance.

 We utilize the also fully equivalent 
Morse-like functions used in the Brenner bond-order potential and which
has been implemented in the PONTIFIX code \cite{Tersoff-Brenner,GaN,SiC}.
\be
V_{ij}^R=\frac{D_0}{S-1} exp(-\beta_{AE} \sqrt{2S}(r-r_0))
\ee
\be
V_{ij}^A=\frac{S D_0}{S-1} exp(-\beta_{AE} \sqrt{2/S}(r-r_0))
\ee
Hence, in fact we fit parameters $S, \beta_{AE}$ keeping $D_0$ and $r_0$ fixed.
The parameters in the angular part are identical with that of Tersoff.
The conversion of the obtained parameters to those used in the Tersoff
potential is straightforward \cite{GaN,SiC,pontifix}.
The required radial Tersoff parameters $A, B, \lambda_1$ and $\lambda_2$ can be
expressed using the Albe-Erhart parameters as follows:
\be
\lambda_1=\beta_{AE} \sqrt{2S}, \lambda_2=\beta_{AE} \sqrt{2/S}.
\ee
\beqn
A=D_0/(S-1)exp(\lambda_1r_0), && \\
B=SD_0/(S-1)exp(\lambda_2r_0),
\eeqn
This conversion must be done when e.g. the LAMMPS code is used.
Example file can be found in the released packages of LAMMPS.

\subsection{The fitting procedure}

 The fitting procedure has been carried out in a few steps.
\\
{\em Step (1)}: First an initial guess of radial parameters have been obtained
using the receipt given by Albe and Erhart \cite{SiC,GaN}.
In particular, $D_0$ and $r_0$ can be estimated from the adhesion energy of gr/Ru(0001) ($E_{adh} \approx 0.2$ eV/C) and equilibrium distance of gr-Ru(0001) ($r_0 \approx 2.1$ $\hbox{\AA}$).
Then using Eqs. (7)-(11) one can estimate the initial guess for $A$,
$B$, $\lambda_1$ and $\lambda_2$.
The proper adjustment of the missing parameters $S$ and $\beta_{AE}$
will also be needed. The details can be found elsewhere \cite{SiC,GaN}.
\\
{\em Step (2)}: Then using the code pontifix \cite{pontifix} 
the entire parameter space has been optimized using a training set
with few general unit cells of weakly bound alloys representing
the bonding of C-Ru alloys.
\\
{\em Step (3)}: Finally, the obtained parameter set
has been further refined by an additional code written in our laboratory
\cite{potfit}.
In this case we consider {\em ab initio} DFT potential energy curves and/or equilibrium
DFT geometries of small gr/Ru(0001) modell systems. 
Using this way of parameter fitting we were able to
get an adequate force field which describe gr/Ru(0001) interfacial bonding properly.
It should be emphasized that steps 1-2 were necessary to obtained
a reliable parameter set. Attempts have been failed when
step 2 has been ignored. 
This could be due to the necessary restriction of the parameters
within a parameter subspace which reflects the properties
of weakly bound cubic alloys.
We find after extensive trials that the adhesion energy and the
morphology is extremely sensitive to the choice of the fitting procedure,
hence the optimization of the most important properties of the
gr/Ru(0001) system ($r_0 \approx 2.1$ $\hbox{\AA}$, $E_{adh} \approx 0.2$ eV/C, hump-and-bump (HAB) morphology, corrugation $\xi \approx 1.0 \pm 0.2$ $\hbox{\AA}$)
is challenging.
E.g. the small change of the parameters can easily alter the HAP topography into
a nanomesh-like one, and/or the adhesion energy can take
extreme values.
Although Step (3) were not essential, already in step (2) we obtained
a more or less reliable parameter set, however, we were able to
further refine the topography by step (3). In particular, the shape of
the protrusions become more hexagonal like and the area of them
is larger which are more similar to the STM results \cite{Batzill}.
Therefore, we used the final refinement step for fine tuning the
force field and make it more adaptive to small modell systems. 

 The traditional way of fitting procedure (see e.g. refs. \cite{SiC,GaN})
does not work in this special case when an interface potential is to be
parameterized.
In a standard situation one should fit the Tersoff function to the
experimental lattice constants, cohesive energies and bulk moduli of various polymorphs
of RuC \cite{RuC}.
However, in this case the interface potential would bind graphene
too strongly to Ru(0001) (chemical adhesion).
Photoelectron
spectroscopy shows that the layer bonding is not carbidic \cite{Parga}.
The bonding situation and the chemical environment is completely
different in gr/Ru(0001) and in RuC.
Even if a weak chemical bonding takes place in gr/Ru(0001), it is far much weaker
than in RuC \cite{DFT:Ru-Hutter,DFT:Ru-Stradi,DFT:Ru_Wang,DFT:Ru_corrug}.
Using a RuC based fitted potential
the adhesion energy of gr/Ru(0001) would be $E_{adh} \gg 1$ eV/C, which is far higher than
the measured and the DFT calculated $E_{adh} \approx 0.2$ eV/C \cite{DFT:Ru-Hutter,DFT:Ru-Stradi,DFT:Ru_Wang,DFT:Ru_corrug}.
Moreover, in the RuC dimer molecule, e.g. the dissociation energy is in the range of
6.6 eV \cite{RuCdim} which is far above the adhesion energy of
gr/Ru(0001) of 0.2 eV/C \cite{DFT:Ru-Hutter}, hence the
dimer properties nor can be used for fitting.
Moreover, the equilibrium distance of the RuC dimer is also too short ($1.66$ $\hbox{\AA}$
when compared with the expected $\sim 2.1$ $\hbox{\AA}$ distance of the
gr/Ru(0001) complex.
Hence one can not use the available experimental data set of RuC for parameterization.

 Therefore one has to choose a different strategy for fitting the Tersoff
parameters.
We find that simple (fictitious) alloys with proper bond distances and cohesive energies  are sufficient for
efficient parameterization.
The training set is necessary only to parameterize
the Tersoff function for reproducing
the weak adhesion energy and bonding distance of C-Ru complexes
at a proper C-C-Ru bonding angle.
In particular, in the fitting data base (training set)
we use the bond distance $r_0=2.1$ $\hbox{\AA}$
and for the adhesion energy $0.2$ eV/C which has been found by
DFT for gr/Ru system \cite{DFT:Ru-Hutter,DFT:Ru-Stradi,DFT:Ru_Wang,DFT:Ru_corrug}.
 In particular,
 the fitting of the angular and radial parts of the interface potential were carried out on a limited data base in step (2)
which includes first neighbor distances and adhesion energy of
fictitious weakly bound dimer, B1 (NaCl), B2 (CsCl) and B3 (ZnS) alloys of C and Ru.

 The obtained parameter set is checked on a trial-and-error basis
on small or medium sized gr/Ru(0001) systems (5000-10000 C atoms on
10 layers Ru(0001)) using MD simulation with the LAMMPS code \cite{lammps}.
Using few steps (in each step with new parameterization procedure
for the angular part and for the other free parameters in the
radial part)  
one can adjust in principle the interfacial Tersoff function
in order to reproduce the most important properties of gr/Ru(0001)
(corrugation, C-Ru bonding energy, adhesion energy, supercell,
proper binding registry).

 Parameters in the angular term ($\gamma$, c,d) are to be determined then by
least square fitting procedures such as installed in the code PONTIFIX \cite{pontifix,GaN,SiC}. 
First, we fit the radial part of the Tersoff function (which is a fully equivalent
Morse-like potential) to the Morse function used for previous simulations.
We use this set of radial parameters as an initial guess for
the parameterization of the full angular-potential.

 In step (3) we used our code for further fitting the parameters
on a more realistic data base.
This final training set include small gr/Ru(0001) systems with
few hundred atoms (with flat gr). The {\em ab initio} DFT potential energy curves
with respect to the variation of the gr-Ru(0001) distance
has been obtained by the D2 \cite{Grimme} and DF2 functionals for van der Waals
interaction using the SIESTA code \cite{siesta}.
Unfortunately, we were unable to get a reasonable
parameter set with the original form of these curves.
The reason became clear after many trials and the study of the literature.
The Grimme's D2 functional seems to overestimate while
the non-local DF2 functional seriously underestimate the adhesion of the gr/Ru(0001) complex.
This leads to too short and too long equilibrium distances for
D2 and DF2, respectively.
We found that our optimized potential as obtained in step (2) gives
a PEC somewhere between the D2 and DF2 potential energy curves (PECs).
In order to keep at least the shape of the PECs for fitting
we shifted upwards the potential energy minima
by 0.2 eV for D2, and by 0.1 eV for DF2.
This somewhat arbitrary modification of the DFT PECs is rationalized
by recent DFT findings in which authors point out similar problems
with these functionals \cite{RPA}.
The more accurate RPA provides much less deep PECs 
than the D2 or DF2 methods for e.g. the gr/Ni(111) system which has
similar adhesion energy to gr/Ru(0001) \cite{RPA}.
Unfortunately no RPA result is available for gr/Ru(0001).
Further details are given in the manuscript.
Using step 3 one can finetune adhesion energy
by adjusting the energy minimum of the PECs included in the
training set (fitting database). In step 2 adhesion can be
adjusted only approximately.
 Using these curves we were able to further refine the PEC obtained in step (2).

 In general, 
the development of a reliable parameter set is required many trial-and-error
simulations besides the optimization of the parameter set (using PONTIFIX).
In each step, after the manual change of the selected parameters ($D_0$, $r_0$, $R_c$, $D_c$)
the rest of the parameters (mostly the angular part) are optimized by PONTIFIX \cite{pontifix}.
With the new parameter set a new MD simulation is done at 0 K in order
to check its performance.
This is repeated many times until the best possible topography is obtained.
The main requirements are the following: the average corrugation is being below $2$ $\hbox{\AA}$ even at 300 K,
moreover, the 0 K structure should be stable at 300 K with
minor corrugation increase,
minimal C-Ru distance $d_{min} > 1.9$ $\hbox{\AA}$,
maximal C-Ru distance $d_{max} < 4.0$ $\hbox{\AA}$,
correct $12 \times 12$ unit supercell,
the adhesion energy is $0.15 < E_{adh} < 0.25$ eV/C,
no decorations occurs on the surface besides the regular shaped humps
(no further protrusions, vacancy islands, or holes).

\begin{table}[t]
\caption[]
{
The fitted Tersoff parameters for the graphene/Ru(0001) interface.
}

\begin{ruledtabular}
\begin{tabular}{lc}
C-Ru (Tersoff, pw)  &      \\
\hline%

A (eV)      & 765.558150 \\
B (eV)      & 74.047131  \\
$\lambda_1$ & 3.838970  \\
$\lambda_2$ & 1.839467      \\
$\gamma$    & 10.080166 \\
c           & 2.497889  \\
d           & 0.212275 \\
h           & 0.211977   \\
$R_c$ ($\hbox{\AA}$)         & 2.976020   \\
$D_c$ ($\hbox{\AA}$)         & 0.561197      \\
$\beta$ ($\hbox{\AA}^{-1}$)  & 2.057177   \\
$\mu$                        & 6.480552        \\
n                            & 1   \\
\hline
\end{tabular}
\end{ruledtabular}
\footnotetext[1]{
The parameters have been fitted to a small data base 
which includes dimer, NaCl (B1) and CsCl (B2) type unit cells
with fictitious bond distances ($d \approx 2.1$, $\hbox{\AA}$).
For fitting the PONTIFIX \cite{pontifix} code has been utilized (step 2)
and subsequently further refined with the code potfit \cite{potfit}
using a small gr/Ru(0001) modell system with flat gr (step 3).
}
\label{T1}
\end{table}